\RequirePackage{ifpdf}
\ifpdf 
\documentclass[pdftex]{sigma}
\else
\documentclass{sigma}
\fi

\def\p{\partial}

\newcommand{\brz}{\{\!\{}
\newcommand{\ktz}{\}\!\}}

\newcommand{\kt}{\rangle}
\newcommand{\br}{\langle}
\newcommand{\bbr}{\br\!\br}
\newcommand{\kkt}{\kt\!\kt}
\newcommand{\pbr}{\prec\!}
\newcommand{\pkt}{\!\succ}

\begin{document}

\allowdisplaybreaks

\renewcommand{\thefootnote}{$\star$}

\renewcommand{\PaperNumber}{018}

\FirstPageHeading

\ShortArticleName{Planarizable Supersymmetric Quantum Toboggans}

\ArticleName{Planarizable Supersymmetric Quantum Toboggans\footnote{This
paper is a contribution to the Proceedings of the Workshop ``Supersymmetric Quantum Mechanics and Spectral Design'' (July 18--30, 2010, Benasque, Spain). The full collection
is available at
\href{http://www.emis.de/journals/SIGMA/SUSYQM2010.html}{http://www.emis.de/journals/SIGMA/SUSYQM2010.html}}}

\Author{Miloslav ZNOJIL}

\AuthorNameForHeading{M. Znojil}

\Address{Nuclear Physics Institute ASCR, 250 68 \v{R}e\v{z}, Czech
Republic}

\Email{\href{mailto:znojil@ujf.cas.cz}{znojil@ujf.cas.cz}}
\URLaddress{\url{http://gemma.ujf.cas.cz/~znojil/}}

\ArticleDates{Received November 30, 2010, in f\/inal form February 21, 2011;  Published online February 25, 2011}

\Abstract{In supersymmetric quantum mechanics the emergence of a
singularity may lead to the breakdown of isospectrality between
partner potentials. One of the regularization recipes is based on a
topologically nontrivial, {\em multisheeted}  complex deformations
of the line of coordinate $x$ giving the so called quantum toboggan
models (QTM). The consistent theoretical background of this recipe
is brief\/ly reviewed. Then, certain supersymmetric QTM pairs are
shown exceptional and reducible to doublets of non-singular ordinary
dif\/ferential equations a.k.a.\ Sturm--Schr\"{o}dinger equations
containing a weighted energy $E \to E  W(x)$ and living in {\em
single} complex plane.}


\Classification{81Q60; 81Q12; 46C15; 81Q10; 34L20; 47A75; 47B50}

\renewcommand{\thefootnote}{\arabic{footnote}}
\setcounter{footnote}{0}

\section{Introduction}

Within the framework of the standard supersymmetric quantum
mechanics (SUSY QM as reviewed, e.g., by Cooper et al.~\cite{Khare})
the interaction potentials $V(x)$  must be assumed {\em regular}.
Otherwise, in a way pointed out by Jevicki and Rodrigues \cite{JR}
the usual correspondence between spectra of the SUSY-related
``partner'' Hamiltonians might disappear. The SUSY-induced pa\-ral\-lels
between the bound states generated by the partner potentials would
break down~\cite{Kharebe}. The main phenomenological merit of the
formalism would be lost. In the language of mathe\-ma\-tics, with the
singularity emerging at least in one of the partner potentials, even
the respective Hilbert spaces may become dif\/ferent. For illustration
of details one may consult, e.g., Fig.~12.1 of~\cite{Khare}
or the explicit construction of an example by Das and Pernice~\cite{DP}.

\looseness=-1
In  the spirit of the ${\cal PT}$-symmetric quantum mechanics (PTS
QM, introduced and made popular mainly in~\cite{BG,BB,pseudo,Ali,BBJ,DDT}) we returned to the Jevicki's
and Rodrigues' problem. In a~comment \cite{NPB} on~\cite{DP} we
recommended the regularization of the strong singularities in $V(x)$
mediated by a suitable {\em small} complex deformation of the line
of coordinates. The requirement of the smallness of the ${\cal
PT}$-symmetric deformation ${\cal C} \to \mathbb{R}$ of the line of
coordinates appeared essential as long as the regularized
Hamiltonians themselves became manifestly non-Hermitian,~$H \neq
H^\dagger$.

During the subsequent development of the theory a full understanding
has been achieved of the formal foundations of the recipe based on
the use of non-Hermitian operators of observables with real spectra.
The recent reviews~\cite{Dorey,Carl,ali,SIGMA} my be recalled as
conf\/irming the full compatibility of ${\cal PT}$-symmetric models
with the standard postulates of quantum mechanics.

The resulting extension of the practical scope of quantum theory
opened the path towards the proposal of a new class of
regularization recipes \cite{uf,tob}. They may be characterized by
the use of {\em arbitrarily deformed} and even {\em topologically
nontrivial} complex curves $x=x(s) \in {\cal C} \subset \mathbb{C}$
of generalized coordinates def\/ined over certain topologically
nontrivial Riemann surfaces ${\cal R}$.

In a number of concrete non-SUSY examples \cite{tobo}, the latter
curves ${\cal C}$ were, typically, chosen as spiralling around the
branch points of the wave functions $\psi(x)$. For this reason we
coined the name of ``quantum toboggan models'' (QTM) for all of the
models supported by a topologically nontrivial (i.e., multisheeted)
integration curve ${\cal C}$ along which the latter wave functions
$\psi(x)$ are, by construction, analytic.

\looseness=-1
In our most recent paper on the QTM \cite{Jak} we returned to~\cite{NPB} and indicated that the menu of possible QTM
regularizations of SUSY may be broader than expected. New and
surprising connections between algebraic (SUSY) and analytic
properties of wave functions were recently revealed~\cite{Jakov}. In
this context the use of the regularizations  leads again to a number
of new open questions (cf., e.g., \cite{Jakovlev}). For this reason
we are also returning to the subject in our present paper.

In addition to an overall outline of the relevant parts of the
theory, we plan to present here also a few new and unpublished
results. First, in Section~\ref{prvni} a compact summary will be
given of the traditional role of SUSY in quantum mechanics. Second,
Section~\ref{druha} will review brief\/ly the introduction and a few
relevant results obtained during the most recent QTM constructions.
Next, Sections~\ref{treti} and~\ref{ctvrta} will of\/fer a thorough
explanation of compatibility between QTM constructions and the
standard probabilistic interpretation and postulates of quantum
mechanics. The key ingredients in these explanations will be the
simplif\/ication of the QTM mathematics via a~``rectif\/ication'' change
of variables and the subsequent construction of the ``standard"
Hilbert space~${\cal H}^{(S)}$ by means of the introduction of
certain not entirely standard inner products.

A few open problems will f\/inally be addressed in Section~\ref{pata}
(including also our main illustrative example of planarization) and
in a few appendices. Thus, in Appendix~\ref{appendixA} we shall explain that the
non-Hermiticity property $H \neq H^\dagger$ of the regularized
Hamiltonians (say, of~\cite{NPB}) is just an artifact of an
inadequate choice of the ``false'' Hilbert space ${\cal H}^{(F)}$
where the superscript may also hint that this space is only ``too
friendly'' (cf.\ also~\cite{SIGMA} for additional details). In
subsequent Appendix~\ref{appendixB} we shall make use of this theoretical
background attaching three Hilbert spaces to the single quantum
system. In Appendix~\ref{appendixC} we shall describe a more concrete
implementation of the resulting innovative ``three-Hilbert-space''
quantum mechanics (THS QM) to the tobogganic version of popular $V
(x)=ix^3$. Finally, all these preliminary considerations will be
fructif\/ied  after return to Section~\ref{pata} where a synthesis of
the idea of SUSY with the idea of ${\cal PT}$-symmetry will be
presented (and summarized in Section~\ref{summary}).

\section{Supersymmetric Schr\"{o}dinger equations}\label{prvni}

Let us recall that, paradoxically, the recent intensive development
of the mathematics of SUSY QM failed to reach its original goals (in
particle physics) but still may be considered very successful,
especially in its role of a tool for study of mutual relations
between dif\/ferent potentials in the quantum bound state problem.

The basic idea of SYSY QM is virtually elementary: in terms of the
two dif\/ferential operators
 \begin{gather}
  A=
 -\frac{d}{dx} + {\cal W}^{}(x) ,\qquad
 B=\frac{d}{dx} + {\cal W}^{}(x)
 \,
 \label{definition}
 \end{gather}
def\/ined in terms of the calligraphic-font ``superpotential'' ${\cal
W}^{}(x)$ one constructs the calligraphic-font partitioned
Hamiltonian
 \begin{gather}
{\cal H}= \left [
 \begin{array}{cc}
H^{(U)}&0\\ 0&H^{(L)}\end{array} \right ]= \left [
 \begin{array}{cc}
  B^{} A^{}
&0\\ 0& A^{} B^{} \end{array} \right ], \label{alice}
 \end{gather}
where the superscripts $^{(U,L)}$ stand for the ``upper'' and
``lower'' partition, respectively. The other two calligraphic-font
partitioned operators are also introduced as playing the role of the
supercharges,
 \begin{gather*}
{\cal Q}=\left [
 \begin{array}{cc}
0&0\\
A^{}&0 \end{array} \right ], \qquad \tilde{\cal Q}=\left [
 \begin{array}{cc}
0& B^{}
\\
0&0 \end{array} \right ] . 
 \end{gather*}
These operators are easily shown to generate one of the simplest
SUSY algebras (cf.~\cite{Khare} for all the details and multiple
consequences).

In the spirit of our comment~\cite{NPB} it is important to consider
operators~(\ref{definition}) without any additional assumption
requiring the reality and regularity of the superpotentials~${\cal
W}^{}(x)$. As a~consequence, operators $A$ and $B$ need not
necessarily play the role of the usual annihilation and creation
operators, etc. For the f\/irst time we used this innovative
f\/lexibility of superpoten\-tials~${\cal W}(x)$ in~\cite{Cannata}
(cf.\ also~\cite{Cannatabe}) where we managed to regularize
 \begin{gather*}
 V \sim \frac{1}{r^2}\ \ \Longrightarrow \ \
V \sim \frac{1}{(t-i\varepsilon)^2}  , \qquad r\in \mathbb{R}^+  ,
\qquad t\in \mathbb{R}
\end{gather*}
via an additional redef\/inition
 \begin{gather}
  A=
 -{\cal T}\frac{d}{dx} + {\cal T}{\cal W}^{}(x) ,\qquad
 B=\frac{d}{dx}{\cal T} + {\cal W}^{}(x){\cal T} .
 \label{redefinition}
\end{gather}
The choice of the new auxiliary involution operator ${\cal T}$
remained unrestricted by the algebraic considerations.

In the above-mentioned amendment of the formalism \cite{Jak} we
merely relaxed the involutivity property ${\cal T}^2=I$ and
postulated
 \begin{gather}
  A=
 -{\cal T}\frac{d}{dx} + {\cal T}{\cal W}^{}(x) ,\qquad
 B=\frac{d}{dx}{\cal T}^{-1} + {\cal W}^{}(x){\cal T}^{-1}
 .
 \label{refinition}
\end{gather}
Now we intend to develop this idea further on.

The f\/irst technical ingredient needed for a consistent build up of
the SUSY version of the theory lies in the necessary clarif\/ication
of the change of the role of the Hermitian conjugation operation
during and after the supersymmetrization. The discussion of this
point will be given below. Here, in introduction, we would only like
to remind the readers that the use of the {\em complex}
superpotentials ${\cal W} = - {\psi'_0}/{\psi_0}$ proved extremely
productive, especially by leading to many new solvable
Schr\"{o}dinger equations (cf.~\cite{MZ2002,MZb2002}) in which the
authors managed to circumvent the obstructions connected with the
``seed'' wave functions $\psi_0$ having nodal zeros.

\section{Tobogganic Schr\"{o}dinger equations}\label{druha}

Even the earliest papers on the models described by
equation~(\ref{equat}) of Appendix~\ref{appendixA} \cite{Caliceti} already reported the
emergence of the real bound-state energies for certain non-real
potentials $V(x)$ and/or for certain non-real paths of integration
${\cal C}(s) \neq \mathbb{R}$ of the Schr\"{o}dinger equation. These
papers found their motivation in perturbation theory and their
authors did not pretend to of\/fer any truly physical predictions. The
situation has changed after the publication of the inf\/luential
letter~\cite{BB}. The subsequent massive return of interest to the
apparently non-Hermitian models with real spectra resulted in the
discoveries of many new exactly solvable analytic models~\cite{ptho}, square-well-shaped models~\cite{ptsqw}, point
interaction models~\cite{Albeverio}, Calogero-type many-body mo\-dels~\cite{Calo} and, last but not least, of the tobogganic models in
which the paths  ${\cal C}(s)$ were chosen as spiralling around the
branch points of their (by assumption, analytic) wave functions~$\psi(x)$.

Teaching by example let us recall here the most elementary (viz.,
force-free, radial, purely kinematical) Hamiltonian~$H$ of~\cite{knot},
 \begin{gather*}
 H =-\frac{d^2}{dx^2}+\frac{\ell(\ell+1)}{x^2} ,\qquad \ell
 \in \mathbb{R} .
\end{gather*}
In a small complex vicinity of the origin any related wave function
$\psi(x)$ has, obviously, the well known form
 \begin{gather*}
 \psi(x) = c_+ \psi^{(+)}(x) +c_- \psi^{(-)}(x),
\end{gather*}
where
 \begin{gather*}
 \psi^{(+)}(x) = x^{\ell+1} + {\rm corrections},\qquad
 \psi^{(-)}(x) = x^{-\ell} + {\rm corrections},
 \qquad |x| \ll 1 .
 \end{gather*}
This means that at a generic real $\ell>-1/2$ the corresponding
Riemann surface ${\cal R}$ will posses a~branch point at $x=0$.
Whenever this value of $\ell$ becomes rational, the corresponding
Riemann surface ${\cal R}$ will be glued of a f\/inite number of
separate sheets. Obviously, the study of the analytic properties of
wave functions only becomes suf\/f\/iciently well known at the integer
values of $\ell$ (cf.\ also a comment on this point in~\cite{knot}).

\subsection{The  single-Riemann-sheet (i.e., cut-plane) models}

The most natural way out of the dif\/f\/iculties which characterize the
general choice of $\ell$ lies, obviously, in the simplest special
choice of $\ell=0$. Even then, the study of the most popular complex
potentials $V(q)=(iq)^{\delta} q^2$ requires a number of technical
simplif\/ications which lead, according to the recommendations in~\cite{BB}, to the most stable results when the path ${\cal
C}(s)$ remains conf\/ined to a single cut complex plane. At any
$\delta\geq 0$ one can then def\/ine an ``optimal'' set of curves
(i.e., $\delta$-dependent asymptotic boundary conditions) for which
{\em all} of the bound-state energy levels $E_n$ behave as smooth
functions of the couplings $\lambda$ or of the other variable
parameters collected in a suitable multiindex~$\vec{\lambda}$
entering the energies as their argument, $E_n=E_n(\vec{\lambda})$.

\begin{figure}[t]
\centering
\includegraphics[angle=270,width=75mm]{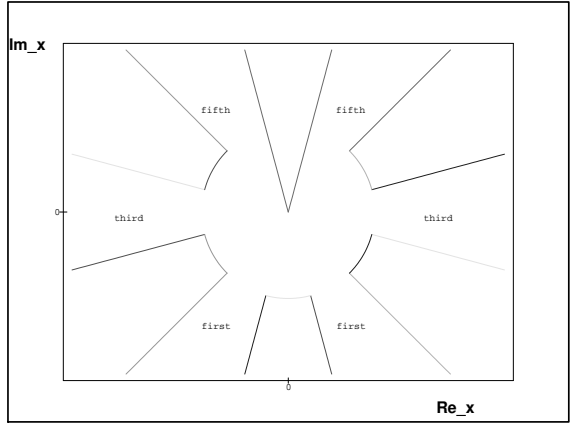}

   \caption{The allowed wedges for the non-tobogganic,
single-complex-plane integration contours for potential
$V(x)=x^2(ix)^8$ and for the preselected wave-function asymptotics
$\psi(x) \sim \exp{-x^6/6}$.
 \label{fdfione}}
\end{figure}

For illustration we may pick up, say, the asymptotically decadic
potential $V(x)=\lambda\, (ix)^8x^2$ and declare, say, the specif\/ic
asymptotics $\psi(x) \sim \exp{(-x^6/6)}$ of wave functions physical
at $|x|\gg 1$. Then, the coupling-independent triplet of the allowed
left-right-symmetric asymptotical wedges exists in a way hinted by
Fig.~\ref{fdfione} where we did set $\lambda=1$ without loss of
generality.

Naturally, beyond the single complex plane (with an upwards-oriented
cut) the heuristically extremely  successful left-right symmetry of
curves ${\cal C}(s)$ might be also preserved~-- this was the basic
idea of the introduction of quantum toboggans in~\cite{tob}.

\subsection{The more-Riemann-sheets models (quantum toboggans)}

Let us assume that our wave functions $\psi(x)$ possess just a
single branch-point singularity (located in the origin) and that a
family of tobogganic curves ${\cal C}^{(\alpha)}(s)$ is def\/ined as
the set of paths over the Riemann surface ${\cal R}$ which are
left-right symmetric in projection to the cut complex plane.
Moreover, we shall assume that these curves remain

\begin{figure}[t]
\centering
\includegraphics[angle=270,width=75mm]{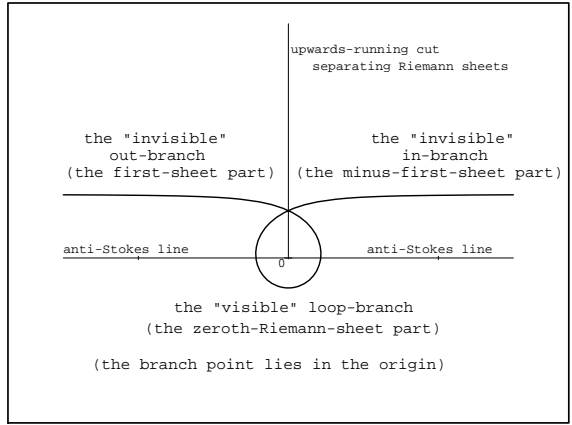}

  \caption{An example of a single-branch tobogganic
curve ${\cal C}^{(\pi)}$.
 \label{fione}}
\end{figure}

\begin{enumerate}\itemsep=0pt

\item[1)]
 locally linear and parallelling the real axis at $|s| \ll 1$; this
means that they may be then prescribed by the formula
 ${\cal C}^{(\alpha)}(s) = -{\rm i}\varepsilon+s +{\cal O}(s^2)$
at a suitable shift-parameter $\varepsilon >0$;

\item[2)]
 turning upwards by the angle $\alpha > 0$ during the growth of
$s \in (0,\infty)$ while disappearing ``behind the cut'' (i.e., to
the f\/irst Riemann sheet, etc.), provided only that $\alpha> \pi/2$;

\item[3)]
 turning upwards by the angle $\alpha > 0$ during the growth of
$t = -s \in (0,\infty)$ while disappearing ``behind the cut'' (i.e.,
to the minus-f\/irst Riemann sheet, etc.), provided only that $\alpha>
\pi/2$;

\item[4)]
 encircling the branch point by the total angle $2\alpha $ and
becoming again approximately linear at large $|s| \gg 1$ (cf.\ the
illustrative sample of ${\cal C}^{(\alpha)}$ with $\alpha=\pi$ given
in Fig.~\ref{fione}).
\end{enumerate}

At the present stage of development of the general theory of quantum
toboggans \cite{confser} our insight in their structure based on the
existing purely numerical studies of their spectra  cannot still be
considered satisfactory. In particular, the bad news were
communicated by Gerd Wessels~\cite{Gerd} who merely managed to show
how the standard numerical methods of Runge--Kutta class succeed in
some of the most standard non-tobogganic models~(\ref{equat}). His
numerical algorithms failed to converge in the domains of the
expected reality of the tobogganic spectra. Similar incomplete
preliminary calculations were also reported by Hynek B\'{\i}la~\cite{Hynben}.

Fortunately, our recent attempted application of perturbation theory
(brief\/ly summarized in Appendix~\ref{appendixC} below) succeeded in the
quantitative estimate of the topology-dependence of the tobogganic
stable-bound-state spectra in the large-$\ell$ dynamical regime. The
calculations have already been performed for the most important
special case of the well known toy model where $V(x) \sim {\rm
i}x^3$. The current $1/\ell$ expansion technique proved applicable
and consistent. Our calculations led to the prediction of the
nontrivial leading-order topology-dependence of the low-lying energy
levels.

These conclusions cannot be directly extended to the domain of small
$\ell$ where the topology-dependence of the tobogganic
stable-bound-state spectra may be expected dif\/ferent. In particular,
it may prove much more sensitive to the details of the behavior of
the wave functions near the branch point. This expectation may be
extracted from the interesting non-tobogganic numerical study~\cite{DDT2}. We believe that a new insight in this technical
subtlety might also prove obtainable after the incorporation of the
very ef\/f\/icient machinery of SUSY QM in combination with the new set
of tricks available within the PTS QM framework. In particular,
solvable examples should be mentioned as an important tool in the
QTM analysis.

\subsection{A quantum-toboggan example which is exactly solvable}

\begin{theorem}[cf.~\cite{knot}]
 The tobogganic Schr\"{o}dinger equation
 \begin{gather}
  -\frac{d^2}{dr^2} \psi (r)+ \frac{\ell(\ell+1)}{r^2}
 \psi (r)= E  \psi (r) ,\qquad
  \ell=n+\frac{D-3}{2}
   \label{ruch}
\end{gather}
which is perturbed by the short-range potential
 $V(r) = \gamma/r^2$ of a quantized strength
 \begin{gather*}
 \gamma=\left (\frac{M}{2N}
 \right )^2-
 \left (n+\frac{D-2}{2}
 \right )^2
\end{gather*}
possesses a closed-form bound-state solution $\psi (r)\in
\mathbb{L}^2({\cal C}^{(\alpha)})$ at
 any integer spatial dimension parameter $D$, integer angular momentum parameter
$n$, integer winding number $N=\alpha/\pi$ and an ``allowed" integer
  $M =  1, 2, 3, \ldots$,
 $
 M \neq 2N, 4N, 6N, \ldots$.
\end{theorem}

\begin{proof} We may set
 $E=\kappa^2$, $z=\kappa r$
 and $\psi(r)=\sqrt{z} \varphi(z)$
with
 \begin{gather*}
 \ell(\ell+1) = \gamma + \left (n+\frac{D-3}{2}
 \right ) \left (n+\frac{D-1}{2}
 \right ) .
\end{gather*}
As long as equation~(\ref{ruch}) is solvable in terms of Bessel
functions,
 \begin{gather*}
 \psi(r) = c_1 \sqrt{r} H^{(1)}_\nu(\kappa r)
 +c_2 \sqrt{r} H^{(2)}_\nu(\kappa r) ,\qquad \nu =
 \ell+1/2
 \end{gather*}
we have to study their asymptotics in the asymptotic (i.e., $\varrho
\gg 1$) wedges
\begin{itemize}\itemsep=0pt

\item ${\cal S}_0 = \{r=-{\rm i} \varrho e^{{\rm
i} \varphi}\,|\,  \varphi \in (-\pi/2,\pi/2)\}$,

\item ${\cal S}_{\pm k} = \{r=-{\rm i} e^{\pm {\rm
i} k \pi} \varrho e^{{\rm i} \varphi}\,|\, \varphi \in
(-\pi/2,\pi/2)\}$, $k \geq 1$.

\end{itemize}

By construction, the tobogganic  contour ${\cal C}^{(\alpha)}$ then
connects the Stokes' wedge ${\cal S}_0$ with the Stokes' wedge $
{\cal S}_{m}$ where $ m=2N$. Thus, we are entitled to use the well
known asymptotic formulae
\begin{gather*}
 \sqrt{\frac{\pi z}{2}} H^{(1)}_\nu(z)
 \exp\left [-{\rm i}\left (z-\frac{\pi(2\nu+1)}{4}
 \right )\right ]=1-\frac{\nu^2-1/4}{2{\rm i} z} + \cdots ,
\\
 \sqrt{\frac{\pi z}{2}} H^{(2)}_\nu(z)
 \exp\left [{\rm i}\left (z-\frac{\pi(2\nu+1)}{4}
 \right )\right ]=1+\frac{\nu^2-1/4}{2{\rm i} z} + \cdots,
\end{gather*}
and conclude that in  ${\cal S}_{2k}$ we may eliminate the
unphysical (i.e., asymptotically non-vanishing) component
$H^{(1)}_\nu(z)$ and get the physical normalizable wave function
proportional to~$H^{(2)}_\nu(z)$. In~${\cal S}_{2k+1}$, on the
contrary, we encounter the physical $H^{(1)}_\nu(z)$ and  unphysical
$H^{(2)}_\nu(z)$. In this way we may start at $s = -\infty$  from
 \begin{gather*}
 \psi^{\rm (left)}(r) = c  \sqrt{r} H^{(2)}_\nu(\kappa r)\qquad ({\rm with}\
 r \in {\cal S}_0)
\end{gather*}
and end up  with $\psi^{\rm (right)}(r)$ given in closed form
 \begin{gather*}
 H^{(2)}_\nu\left (ze^{{\rm i}m\pi}\right )=
 \frac{\sin (1+m)\pi\nu}{\sin \pi \nu}
 H^{(2)}_\nu(z)+
 e^{{\rm i}\pi\nu}
 \frac{\sin m\pi\nu}{\sin \pi \nu}
 H^{(1)}_\nu(z) .
 \end{gather*}
Obviously, the latter function will vanish at $s = +\infty$ for the
quantized  angular momenta $\nu$:
 \begin{gather*}
  m\nu = {\rm integer},\qquad \nu \neq {\rm integer} \quad
  \Longrightarrow \quad \ell =\frac{M-N}{2N} .\tag*{\qed}
\end{gather*}
\renewcommand{\qed}{}
  \end{proof}

From a purely pragmatic point of view the weakest point of the
above, methodically important exactly solvable model of~\cite{knot} lies in the complete absence of a conf\/ining
external potential. An instability of this state with respect to
random perturbations ref\/lects just its being embedded in the
continuum of scattering states. Unfortunately, the incorporation of
any asymptotically conf\/ining force would already require the
numerical solution of the corresponding tobogganic Schr\"{o}dinger
equation.

\begin{figure}[t]
\centering
\includegraphics[angle=90,width=75mm]{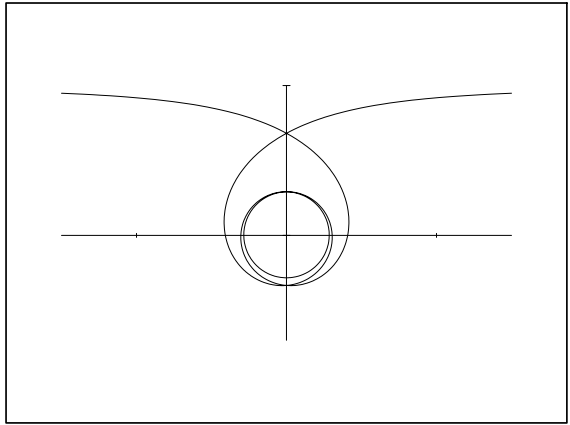}

  \caption{A tobogganic curve which three times
encircles the branchpoint. \label{fiongge}}
\end{figure}

\section[The Sturm-Schr\"{o}dinger ``canonical" form of the toboggans]{The Sturm--Schr\"{o}dinger ``canonical" form of the toboggans} \label{treti}

Once we def\/ined quantum toboggans as ordinary dif\/ferential
Schr\"{o}dinger equations integrated over paths ${\cal C}(s)$
connecting several Riemann sheets of the wave functions $\psi(x)$,
it is natural to complement, in the next step of the analysis, the
single-loop tobogganic-curve of Fig.~\ref{fione} by its
multiply-spiralling descendants sampled, say, in Fig.~\ref{fiongge}.
At the same time, in our present paper we shall not move to the next
step of generalization in which more branch points are admitted in
${\cal R}$ (in this respect, e.g.,  \cite{twobr} may be
consulted).

For all of the left-right symmetric central spirals one can consider
the {\em tobogganic} ordinary dif\/ferential QTM Schr\"{o}dinger
equation
 \begin{gather*}
 H^{(1)}\,\psi^{(1)}_n(r)
 = E_n\,
 \psi^{(1)}_n(r) ,
 \qquad
 H^{(1)}=-\frac{d^2}{dr^2}
 +V^{(1)}(r)
 \end{gather*}
and introduce a change of  variables (say, $r \to y$) which is, up
to some constants, such that $r \sim y^\alpha$ with a suf\/f\/iciently
large exponent $\alpha \in (1,\infty)$. Under such a mapping the two
radial rays in the initial, tobogganic complex $r$-plane will be
mapped on the two radial rays in the complex $y$-plane {\em with a
smaller angle} between them. Thus, any angle between the radial rays
becomes diminished by the mapping  even when the initial rays are
located on a multisheeted Riemann surface ${\cal R}$ of variable~$r$. In this manner some of the rays which were originally hidden
behind the cut (i.e., which belonged to the invisible Riemann sheets
in variable~$r$) will eventually move to the zeroth Riemann sheet in
the new coordinate~$y$, i.e., they become visible in the standard
complex plane of~$y$, say, with an upwards-running cut.

The detailed description of this technique which will rectify (i.e.,
straighten up) the QTM curves sampled in Figs.~\ref{fione} and~\ref{fiongge} (i.e., lower the total angle between their
asymptotics) may be found elsewhere~\cite{confser}. Here, let us
only mention that in the literature one quite often f\/inds the
alternative, de-rectifying complex mappings $x \sim y^\beta$ with a
small exponent $\beta \in (0,1)$. On Riemann surfaces this mapping
acts in the opposite direction. Most often, it is used just to
transform the exactly solvable harmonic oscillator into the exactly
solvable bound-state part of the Coulombic (or Kepler) quantum
system~\cite{Fluegge}. In the latter setting, both the initial and
f\/inal coordinates remain real. The resulting, new Schr\"{o}dinger
equation for bound states acquires the form of the Sturmian
eigenvalue problems where the energy appears multiplied by a
non-constant (and, in our present paper, non-calligraphic-font)
weight function~$W(y)$,
 \begin{gather*}
 H^{(2)} \psi^{(2)}_n(y)
 = E_n W(y)
 \psi^{(2)}_n(y)
 ,
 \qquad
 H^{(2)}=-\frac{d^2}{dy^2}
 +V^{(2)}(y) .
 \end{gather*}
In our present QTM setting we shall use $\alpha>1$ and emphasize
that both the initial $r$ and f\/inal $y$ are complex in general. This
means that in the respective usual and friendly or ``false'' Hilbert
spaces ${\cal  H}_{(1,2)}^{(F)}=\mathbb{L}_2({\cal C})$
we have to deal with the manifestly non-self-adjoint operators,
 \begin{gather}
 W^{} \neq
 \left [ W^{}
 \right ]^\dagger\qquad {\rm and }\qquad
 H^{(1,2)} \neq
 \big [ H^{(1,2)}
 \big]^\dagger\qquad
   {\rm in}\qquad   {\cal
 H}_{}^{(F)}={\cal
 H}_{(1,2)}^{(F)} .
  \label{haSErovni}
 \end{gather}
Just for illustration it is easy to verify that the change of
variables
 \begin{gather*}
  \mathrm{i} x=(\mathrm{i} z)^2 ,\qquad \psi
_n(x)=\sqrt{z} \varphi _n(z)  
\end{gather*}
with $\alpha=2$ returns  us from the f\/irst nontrivial tobogganic
form of the standard harmonic-oscillator Schr\"{o}dinger equation to
a \emph{non-tobogganic} sextic-oscillator Sturm--Schr\"{o}dinger
equation $H\varphi=EW\varphi$ which remains manifestly
$\mathcal{PT}$-symmetric and def\/ined on an U-shaped contour where~$W
> 0$ \cite{SieglCoul},
 \begin{gather*}
 \left(-\frac{d^2}{dz^2}+4z^6+\frac{4\alpha^2-1/4}{z^2}
\right) \varphi_n(z)=-4E_nz^2\varphi_n(z) .
 \end{gather*}
In general cases the rectif\/ication strategy is often further
simplif\/ied and reduced just to the use of the winding-numbered paths
of both the $r$- or $y$-implementations
 \begin{gather*}
 {\cal C}^{(N)}(s)=-{\rm i}\left [{\rm i}(s-{\rm i}\varepsilon)
 \right ]^{2N+1} ,\qquad
  s \in (-\infty,\infty)
\end{gather*}
with integer $N$. More details concerning the concrete realizations
of this technique should be sought elsewhere (e.g., in our
review paper~\cite{confser}).

\section[Towards the non-Dirac metrics $\Theta\neq I$ in the physical Hilbert spaces of states]{Towards the non-Dirac metrics $\boldsymbol{\Theta\neq I}$\\ in the physical Hilbert spaces of states}\label{ctvrta}

\subsection[The non-Sturmian constructions with $W= I$]{The non-Sturmian constructions with $\boldsymbol{W= I}$}

The very brief account of the quantum theory using the non-standard
(often called non-Di\-rac~\cite{Carl}) assumption~(\ref{haSErovni})
may be found in~\cite{SIGMA} or, in an even more compressed
form, in Appendix~\ref{appendixA} below. Its historical roots are in fact very
recent since the explicit quantum-theoretical interpretation of
${\cal PT}$-symmetric models as independently revealed by several
groups of authors only dates back to the years 2001 and 2002~\cite{pseudo,Ali,BBJ}.

It is necessary to add that the basic idea of the unphysical nature
of the ``friendly but false'' Hilbert spaces ${\cal H}^{(F)}$ (with
the trivial, often called Dirac's metric $\Theta^{(F)}=I$) and of
the unitary equivalence between the alternative physical
realizations  ${\cal H}^{(S)}$ (with $\Theta^{(S)}=\Theta \neq I$)
and  ${\cal H}^{(P)}$ (with unfriendly Hamiltonians $\mathfrak{h}$
but still trivial $\Theta^{(P)}=I$) of the space of states already
appeared in nuclear physics before 1992~\cite{Geyer}.

Another marginal addendum to Appendix~\ref{appendixA} concerns notation: one has
to recollect  that the physical Hilbert space ${\cal H}^{(S)}$ is
def\/ined as equipped with an amended Hermitian-conjugation operation~(\ref{adden}) which co-applies also to the operators in the form
 \begin{gather*}
 {\cal A} \to {\cal A}^\ddagger:=\Theta^{-1}
 {\cal A}^\dagger \Theta
 \end{gather*}
such that we may deduce the reality of the spectrum of $H$ by its
(crypto-)Hermiticity $H = H^\ddagger$ realized as ``hidden'' in space
${\cal H}^{(S)}$. Usually one proceeds in an opposite direction.
Keeping in mind the def\/inition of the conjugation ${\cal
A}^\ddagger$ the main problem usually lies in the necessary
construction of the metric operator~$\Theta=\Theta^\dagger> I$.

As long as there exist many alternative operators~$\Theta=\Theta(H)$
for a given Hamiltonian, dif\/ferent versions of the correct
interpretation of the quantum system in question prove obtainable in
general. In more detail, the situation is summarized in Appendix~\ref{appendixB}
below.

\subsection[The transition to $W \neq I$]{The transition to $\boldsymbol{W \neq I}$}

There exist a few not too essential though still nontrivial
dif\/ferences between the non-tobogganic and tobogganic quantum
systems. They may formally be reduced to the dif\/ferences between the
Sch\"{o}dinger and Sturm--Schr\"{o}dinger time-independent equations
where $W=I$ and $W \neq I$, respectively.

We now have to clarify the possibility and methods of construction
of the metrics $\Theta$ for toboggans, say, in their rectif\/ied
Sturm--Schr\"{o}dinger representation. This representation might also
be called ``planarized'' (i.e., def\/ined in a single complex plane,
admissibly with cuts), but we shall rather reserve the latter term
for the narrower families of the models in which {\em all} the
Riemann surface ${\cal R}$ is mapped on the single complex plane
with a single cut or, alternatively, without any cuts at all.

Naturally, the latter specif\/ication of the concept of the
``planarizability'' would be much narrower, requiring that there is
just a f\/inite number of sheets forming the planarizable Riemann
surface ${\cal R}$ in question. Even in such a form its use may
still prove overcomplicated from the purely pragmatic point of view.
For this reason, in what follows, we shall unify and further reduce
the two alternatives and speak about the planarization-mapping
correspondence ${\cal R}\leftrightarrow \mathbb{C}$ if and only if
the mapping itself is given or known in an explicit form of some
well-def\/ined change of the variables.

The importance of distinguishing between the planarizable and
non-planarizable QTM has not yet been addressed in the literature.
This motivated also our present considerations in which we intend to
emphasize that the concept of the planarizable Riemann surfaces is
in fact not too robust (a small change of interaction may make them
non-planarizable). At the same time, the corresponding expected
enhanced sensitivity of the QTM spectra, say, to the small
variations of the angular momenta~$\ell$ is not too surprising.
Indeed, the similar strong-sensitivity phenomenon has already been
observed even in non-tobogganic models where it was quantitatively
described, e.g., by Dorey et al.~\cite{DDT2}. Unfortunately, its
quantitative study in QTM context seems much more dif\/f\/icult. In this
sense our present paper of\/fers just a few f\/irst steps in this
direction.

Our f\/irst message is that the key idea (attributed to Dyson~\cite{Geyer}) of the use of the ``non-unitary Fourier
transformation''~$\Omega$ and of the corresponding def\/inition
$\Theta=\Omega^\dagger\Omega$ of the metric proves transferrable to
the QTM context in a rather straightforward manner.

In the more technical terms one has to start again from the doublet
of Sturm--Schr\"{o}dinger equations
 \begin{gather*}
 H^{}\,|\,{\lambda}\rangle
 =\lambda\,W\,|\,{\lambda}\rangle
,\qquad
 H^\dagger \,|\,{\lambda'}\rangle\!\rangle
 =\lambda'\,W^\dagger\,|\,{\lambda'}\rangle\!\rangle,
\end{gather*}
where $\lambda$ denotes any element in the (by assumption, shared)
spectrum of the ``conjugate Hamiltonians''~$H$ and $H^\dagger$.

\looseness=-1
Under another assumption of the reality and bound-state
form\footnote{I.e., discrete and bounded from below.} of this
spectrum the formulation of the general theory may again proceed in
full analogy with the non-Sturmian and non-tobogganic cases where
$W=I$. In particular, besides the point spectrum (eigenvalues), due
attention must also be paid to the possibility of the existence of
the continuous spectrum as well as to the proofs of the absence of
the residual part of the spectrum. Still, as long as our operators
are mostly just the linear dif\/ferential operators of second order,
the rigorous discussion of these questions (plus of the questions of
the domains, etc) remains as routine as in the non-tobogganic cases.

\subsection[Dyson-type mappings $\Omega\neq \Omega^\dagger$ for $W \neq I$]{Dyson-type mappings $\boldsymbol{\Omega\neq \Omega^\dagger}$ for $\boldsymbol{W \neq I}$} \label{fff.}

Teaching by example let us return to the notation conventions of~\cite{SIGMA} and write
 \begin{gather*}
 |\psi \pkt = \Omega\,|\psi \kt ,
 \qquad
 |\psi \kt \in {\cal V} ,
 \qquad
 |\psi \pkt \in {\cal A}
 \end{gather*}
for ket vectors in ${\cal H}^{(P)}$, taking into consideration also
their duals,
 \begin{gather*}
  \pbr\,\psi\,| = \langle\,\psi\,|\,
 \Omega^\dagger \in
 {\cal A}'
 \end{gather*}
as well as the Hamiltonians $h = \Omega H \Omega^{-1}$ and the
weight operators $w = \Omega W \Omega^{-1}$ def\/ined as acting in
${\cal H}^{(P)}$.

In this language the doublet of the Dieudonn\'e-type generalized
constraints
 \begin{gather*}
 h^\dagger = \left (\Omega^{-1}
 \right )^\dagger H^\dagger \Omega^\dagger= h ,
 \end{gather*}
 and
 \begin{gather*}
  w^\dagger = \left (\Omega^{-1}
 \right )^\dagger W^\dagger \Omega^\dagger= w ,
 \end{gather*}
implies that we get, after a trivial re-arrangement, the pair of
equations
 \begin{gather*}
 H^\dagger = \Theta H \Theta^{-1} ,
 \qquad
 W^\dagger = \Theta W \Theta^{-1} ,\qquad \Theta =
 \Omega^\dagger \Omega .
 \label{prope}
 \end{gather*}
They are def\/ined directly in the unphysical and auxiliary but,
presumably, maximally com\-pu\-ta\-tion-friendly space~${\cal H}^{(F)}$.

The further development of the theory is more or less obvious~-- a
few further technical details are made available in Appendix~\ref{appendixB}
below.

\section{The planarized  toboggans} \label{pata}

In the light of the results described in Appendix~\ref{appendixC} the physical
source of the reliability of the large-$\ell$ spectral estimates of~\cite{anom} might be seen in the large magnitude of the
distance (i.e., of the length of the path on which the dif\/ferential
expression is def\/ined) between the branch-point singularity and the
position of the minimum/minima of the related ef\/fective potential.

We now intend to emphasize that the interaction-dependent
multiplicity of the latter mi\-ni\-ma~\cite{Omar} together with the
comparability of their respective distances from the branch point
indicates that one must be very careful with any {\it a priori}
extrapolation of any non-QTM observation to its possible QTM
analogues.

The latter warning led us also to our present proposal of
restricting one's attention solely to the models in which one could
eliminate the multisheeted form of ${\cal R}$ by the purely analytic
means of the the above-discussed rectif\/ication transformation.

As long as we are dealing here just with the models possessing a
mere single branch-point singularity we may work with the necessary
rectif\/ication changes of variables in closed form. As a consequence,
we intend to achieve a maximal prof\/it from the maximalized
simplicity of the replacement of the tobogganic version
$H |\psi\kt=E |\psi\kt$ of Schr\"{o}dinger equation by its
single-plane Sturm--Schr\"{o}dinger eigenvalue counterpart
$H |\psi\kt=E W |\psi\kt$ containing its energy in the term
with a weight factor $W\neq I$. In other words, we shall restrict
our choice of dynamics by the formal requirement that the latter
equation does not contain any strong singularities anymore.

On this background we shall return to the idea of the possible QTM
implementations of the supersymmetric partnership which has been
originally discouraged by our recent general study~\cite{Jak} where
we did not assume any planarizability of the  QTM Riemann sheet~${\cal R}$.

\subsection{Asymptotically harmonic oscillator as
 the simplest planarizable toboggan}

\begin{figure}[t]
\centering
\includegraphics[angle=270,width=75mm]{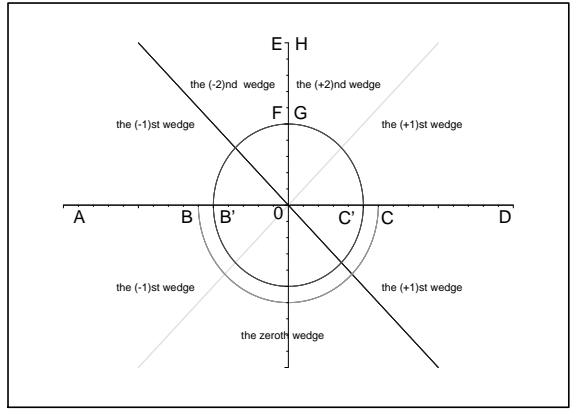}

 \caption{The typical harmonic-oscillator integration
curves visible on the f\/irst Riemann sheet of complex~$x$ (see the
text  for a detailed explanation). The upwards-running cut starts at
the origin; arbitrary units. \label{freon1}}
\end{figure}

\begin{figure}[t]
\centering
\includegraphics[angle=270,width=75mm]{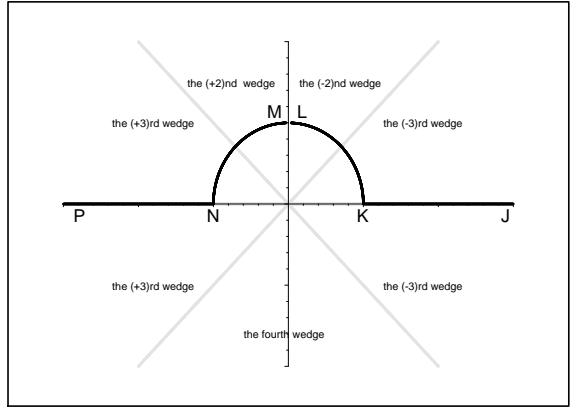}

 \caption{The beginning (J-K-L) and the end (M-N-P) of
the unrectif\/ied tobogganic curve lying on the second Riemann sheet.
The upwards-running cut starts at the origin; arbitrary units. \label{freon2}}
\end{figure}

Let us assume that our potential is asymptotically quadratic, i.e.,
$V(x) = x^2+{\cal O}(x)$ at the large complex $x$. Then the most
general form of the wave function $\psi(x) \sim c_+ e^{x^2/2} +c_-
e^{-x^2/2}$ asymptotically vanishes (and ``remains physical") either
at $c_+=0$ (this takes place either in the $(\pm 1)$st Stokes wedges
of Fig.~\ref{freon1} or in the $(\pm 3)$rd Stokes wedges of
Fig.~\ref{freon2}) or at $c_-=0$ (this takes place either in the
zeroth and $(\pm 2)$nd Stokes wedges of Figs.~\ref{freon1} and~\ref{freon2} or in the last, $(+ 4)$th = $(- 4)$th Stokes wedge of
Fig.~\ref{freon2}).

Let us assume, in addition, that the Riemann surface ${\cal R}$ of
analyticity of our eligible physical wave functions $\psi^{(+)}(x) =
e^{+x^2/2 + {\cal O}(x)}$ or $\psi^{(-)}(x) = e^{-x^2/2 + {\cal
O}(x)}$ is just two-sheeted, i.e., that near the origin we have
$\psi(x) \sim x^{J/2}$ with an odd integer $J$.

\begin{figure}[t]
\centering
\includegraphics[angle=270,width=75mm]{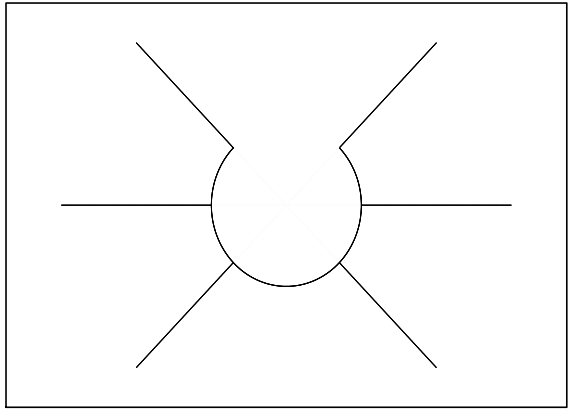}

  \caption{Complex $y$-plane and the planarized triplet
of $x$-paths of Figs.~\ref{freon1} and~\ref{freon2}. \label{freone}}
\end{figure}

For the sake of def\/initeness, we may pick up angular momentum
$\ell=1/2$ giving $J=3$. Then, the elementary change of variables
$ix = (iy)^{2}$ (with full details either easily derived or taken,
say, from~\cite{confser}) gives the new, equivalent form of
wave function $\varphi(y)$ which is, by our assumption, regular or
at most weakly singular in the origin. This means that the change of
variables $x \to y$ maps the two-sheeted Riemann surface\footnote{Shown in Figs.~\ref{freon1} and \ref{freon2} and
pertaining to our ``initial'', tobogganic as well as non-tobogganic,
asymptotically harmonic-oscillator-like wave functions $\psi(x)$.} ${\cal
R}$
onto the single `target'' complex plane $\mathbb{C}$ of
Fig.~\ref{freone} which, formally, plays the role of a
single-sheeted Riemann surface of the new, ``Sturmianic''~\cite{Hendrik} wave function $\varphi(y)$.

By construction of our particular illustrative example, there are no
singularities and no cuts in the latter complex plane. For the sake
of simplicity the boundaries of the Stokes wedges were also omitted in Fig.~\ref{freone} since they only prove relevant
in the asymptotic domain where $|y| \to \infty$.

\subsection{The planarizable supersymmetric toboggans} \label{sesta}

Once we return, in a climax of our preceding QTM-related exposition
and considerations, to the problem of the possible SUSY partnership
between strongly singular potentials, we may now feel conf\/ident that
the formalism is prepared for the rigorous regularization treatment
of the changes of the strength of the singularities in the
potentials.

For the sake of def\/initeness let us start from
equation~(\ref{definition}) and assume that the generic analytic wave
functions $\psi(r)$ of the quantum system in question exhibit a
general power-law behavior in the origin,
 \begin{gather*}
 \psi_0^{(\gamma)}(r) \approx r^{\gamma+1/2},\qquad |r| \ll
 1 ,
 \qquad \gamma \in \mathbb{R} .
 \end{gather*}
This implies that the related Riemann surface ${\cal R}$ will be
multisheeted in general, and $K$-sheeted (with $K=K(\gamma)<\infty$)
for special $\gamma = $ rational. This is one of the facts which
make the problem of QTM spectra rather nontrivial in general, and
this is also one of the reasons why we decided to pay here attention
to the models with f\/inite $K(\gamma)$. These models f\/ind a natural
rectif\/ication-transformation regularization which eliminates the
branch point(s) and ``planarizes'' the model in the way explained
above.

In any case, our ``calligraphic-font'' superpotential of
equation~(\ref{definition}) acquires now one of the most elementary
singular forms,
 \begin{gather}
  {\cal W}_{}(r) ={\cal W}_{}^{(\gamma)}(r) = - \frac{\p_r
 \psi_0^{(\gamma)}(r)
 }{
 \psi_0^{(\gamma)}(r)
 }= \chi(r)-\frac{\gamma+1/2}{r},
 \label{N}
 \end{gather}
where $\chi(r)$ may be an arbitrary function which is suf\/f\/iciently
regular in the origin. Thus, we know the doublet of operators
$A=\p_q+{\cal W}$ and $B=-\p_q+{\cal W}$ and easily deduce the
explicit form of the ``upper'' and ``lower'' sub-Hamiltonians in
equation~(\ref{alice}),
 \begin{gather*}
 H^{(U)}=B\cdot A=\hat{p}^2+{\cal W}^2-{\cal W}'
 , \qquad
 H^{(L)}=A \cdot B=\hat{p}^2+{\cal W}^2+{\cal W}' .
\end{gather*}
Near the origin, their dominant parts read as follows,
 \begin{gather*}
 H^{(U)}=\hat{p}^2+\frac{\gamma^2-1/4}{r^2} + {\cal O}\left (\frac{1}{r}
 \right )
 , \qquad
 H^{(L)}=\hat{p}^2+\frac{(\gamma+1)^2-1/4}{r^2} + {\cal O}\left (\frac{1}{r}
 \right ) .
 \end{gather*}
We see that the SUSY correspondence between operators $H^{(U)}$ and
$H^{(L)}$ leaves our auxiliary integer quantity $K(\gamma)$
invariant. This implies that in any QTM version (i.e., independently
of the winding number~$N$), both of the SUSY-partner QTM operators
may be made regular and planar (=planarized) {\em simultaneously}.

\begin{figure}[t]
\centering
\includegraphics[angle=270,width=75mm]{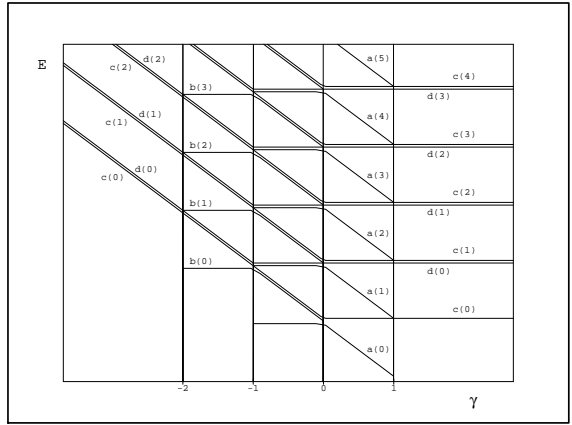}

\caption{The $\gamma$-dependence of the spectrum of
the supersymmetric model which is generated by the singular
superpotential~(\ref{N}) where $\chi(r)=r$. \label{reone}}
\end{figure}

An important remark should be added here since for the real $\gamma$
we have $H^{(U)}$ and  $H^{(L)}$ which depend on the mere absolute
values of $|\gamma|$ and  $|\gamma+1|$, respectively. This implies
the emergence of various interesting structures and asymmetries in
the spectra. Some of them were described,  in~\cite{MZ2002},
via the exactly solvable model with linear $\chi(r) = r$. Some of
the other features of this harmonic-oscillator-like solvable toy
model were discussed in proceedings \cite{srni}. Unfortunately, due
to an error in printing, the illustrative spectrum did not appear in
the paper. Fortunately, this information has not lost its appeal
with time so we now present it  via Fig.~\ref{reone} at last. What
we see in this picture are
\begin{enumerate}\itemsep=0pt

\item[1)] the ``far left'' energies which are completely degenerate,
 \begin{gather*}
E^{(U)}_N \ [\equiv c(N)] = E^{(L)}_N \ [\equiv d(N)],\qquad
\gamma \in (-\infty,-2) ;
\end{gather*}

\item[2)] the ``near left'' energy levels with $\gamma \in (-2,-1)$, exhibiting
a Jevicki--Rodrigues-like breakdown of SUSY~\cite{JR} since the new
levels $ E^{(L)}_N$ $[\equiv b(N)]$ remain non-degenerate;

\item[3)] the ``central-domain'' levels
with $\gamma \in (-1,0)$,
 \begin{gather*}
 E^{(U)}_N \ \! [\equiv a(N)] <
 E^{(L)}_N \ \! [\equiv b(N)] =
 E^{(U)}_N \ \! [\equiv c(N)] <
 E^{(L)}_N \ \! [\equiv d(N)] = a(N+1) < \cdots.
\end{gather*}
Note that one obtains the usual harmonic oscillator at $\gamma =
-1/2$, plus its smooth extension which ceased to be equidistant~--
still, the SUSY is unbroken;

\item[4)] the ``near right'' domain of $\gamma \in (0,1)$ where
the series  $ E^{(L)}_N$ $[\equiv b(N)]$ ceases to exist;

\item[5)] the ``far right'' spectrum
with $\gamma \in (1, \infty)$ and degeneracy
\begin{gather*}
 E_{(L)}^{(+\alpha)} \ [\equiv c(N)] = d(N-1), \qquad
 N > 0.
\end{gather*}
\end{enumerate}

Another concluding remark should be made on the
complex-conjugation-mimicking ope\-ra\-tors~${\cal T}$ as introduced in~\cite{Jak}. Their role has been recalled here via
equations~(\ref{redefinition}) and (\ref{refinition}). Obviously, in a
planarized representation of any QTM and/or SUSY QTM system, these
ope\-ra\-tors will mediate certain complex rotations by angles $\varphi
= 2\pi/K(\gamma)$. This means that they will operate as mutual
mappings between the neighboring rectif\/ied paths of the complex
coordinates $y=\hat{\cal C}(s)$ in the {\em single-sheeted}
$y$-plane without any cuts.

\begin{figure}[t]
\centering
\includegraphics[angle=270,width=75mm]{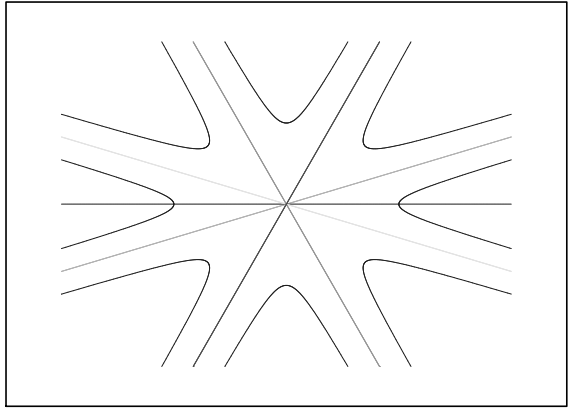}

 \caption{A sample of the planarized system of the
eight QTM trajectories of $y \in {\cal T}^n\hat{\cal C}(s)$. \label{fsjne}}
\end{figure}

The sequence of the individual maps ${\cal T}\hat{\cal C}(s)$,
${\cal T}^2\hat{\cal C}(s)$, ${\cal T}^3\hat{\cal C}(s)$, $\ldots$
may be represented, schematically, by the hyperbolas in
Fig.~\ref{fsjne} where we assumed that the whole QTM system has
been planarized (i.e., there is no cut). Obviously, this picture
strongly suggests that the quantitative analysis of the spectra will
probably exhibit a growing dif\/f\/iculty with the growth of the index
$K(\gamma)= 0, 1, \ldots$ which characterizes a ``hidden'' topology
and symmetry of the QTM system.

\section{Conclusions} \label{summary}

In the mathematical language of operator theory on can say that the
QTM (or, if you wish, also ${\cal PT}$-symmetric and other)
Hamiltonians are similar to Hermitian operators \cite{Dieudonne}. In
the literature one might f\/ind a lot of rigorous results in this
direction ({\it pars pro toto} let us mention the brief note by
Langer and Tretter~\cite{LT} from~2004).

In the equally rigorous language of integrable models, the knowledge
and an appropriate reinterpretation of the  IM/ODE correspondence
did even provide one of the most valuable proofs of reality of the
spectra from $H \neq H^\dagger$ (cf.~\cite{DDT}).

Incidentally, the similar though much easier (and, hence, less often
cited) proofs of the reality of spectra also emerged in the domain
of the (new) exactly solvable models~--  systematic study~\cite{Geza}
may be read as a collection of examples with further references.

On a purely heuristic level one of the most ef\/f\/icient formal
features of the candidates $H \neq H^\dagger$ for Hamiltonians must
be seen in their ${\cal PT}$-symmetry, i.e., in the peculiar
intertwining property $H^\dagger {\cal P} = {\cal P}\,H$ where
${\cal P}$ is, unexpectedly often, an operator of (spatial) parity.

One should not forget that the boom of interest in ${\cal
PT}$-symmetric models grew even beyond the boundaries of quantum
theory, viz, to various  non-quantum  applications of toy models $H
\neq H^\dagger$ in magnetohydrodynamics (cf.\ G\"{u}nther et al.~\cite{uwe}) or in optics (R\"{u}ter et al.~\cite{nature}) etc.

In a parallel recent history of development of the purely pragmatic
use of the concepts of supersymmetry in Quantum Mechanics (cf.,
e.g.,~\cite{MZ2002}) one may notice the existence of the regular
turns of emphasis from the appeal of the underlying algebra of
operators to the necessity of studying the spaces of wave functions
and the related boundary conditions. In this sense we consider the
possibility of construction of SUSY QM models as systems living on
topologically nontrivial Riemann surfaces~\cite{Jak} most appealing.

In our present considerations we used this formalism for addressing,
directly, the family of problems related to the names of Jevicki and
Rodrigues~\cite{JR}. As long as these problems emerge, as a rule, in
connection with the emergence of singular interaction potentials,
their potential relevance in physics is undeniable. Nevertheless,
our present text also emphasized the presence of many technical
obstacles in this ef\/fort.

In summary, the present innovative use of the complexif\/ied
coordinates~$x$ (or~$y$) and of the non-standard, non-Hermitian
operators in SUSY context may be expected to throw new light on the
multiple connections between the abstract algebras (sampled by SUSY)
and nonlinear symmetries (sampled here by~${\cal PT}$-symmetry),
especially in their concrete transparent representations in terms of
the ordinary dif\/ferential operators.

Needless to add that our text still left multiple questions entirely
open. Still, some of these questions (ranging from the abstract
theories down to very concrete estimates and calculations) cannot
remain unanswered  in a long perspective, provided that we wish to
employ the proposed concepts in a more concrete phenomenological
context. For this purpose we expect that the future research will
move towards a deeper synthesis of the algebraic study of the
foundations of SUSY QM with the truly analytic-function approaches
to the mathematical as well as physical puzzles connected with the
models living on more than one sheet of a Riemann surface of the
(presumably, analytic) wave functions.


\appendix

\section{Hilbert spaces endowed with nontrivial inner products} \label{appendixA}

Among applications of quantum theory one only rarely encounters
unusual models like, e.g., the quantum clock in which the quantum
observable is the time \cite{Hoo}. The wealth of possibilities
of\/fered by the abstract formalism is disproportionately often
reduced just to the description of the one-dimensional motion of a
point (quasi)particle. The measurement is then understood to be the
measurement of position $x$ while the evolution of the wave function
$\psi(x)$ in time is assumed controlled (i.e., generated) by the
Hamiltonian $H=-d^2/dx^2+V(x)$ where $H_0=-d^2/dx^2$ represents
kinetic energy and where a real function $V(x)$ mimics the ef\/fects
of an external medium.

In such a context Bender and Boettcher \cite{BB} evoked a lot of
discussions by considering a one-parametric set of purely imaginary
functions $V^{(\delta)}(x)=x^2 ({\rm i}x)^\delta$ with arbitrary
real $\delta \geq 0$ and by disentangling, simultaneously, any
connection between the measurable values of position $q \in
\mathbb{R}$ and the admissible values of the variable $x \in {\cal
C}(s)$ in $\psi(x)$. Here, the real parameter $s \in \mathbb{R}$
just parametrized an {\it ad hoc} complex curve such that ${\cal
C}(s) \neq \mathbb{R}$ in general (cf., e.g., review paper~\cite{Carl} for more details).

In such a setting the concept of quantum toboggans has been
introduced by the mere replacement of condition ${\cal C}(s) \in
\mathbb{C}$ by condition  ${\cal C}(s) \in {\cal R}$ where symbol
${\cal R}$ represents a nontrivial, multisheeted Riemann surface
pertaining to wave functions $\psi(x)$. In this case the curves
${\cal C}(s)$ may be spiral-shaped (hence the name), and only their
topology must be compatible with the domains of analyticity of the
underlying time-independent Schr\"{o}dinger equation
 \begin{gather}
 H \psi(x) = E \psi(x) ,
 \qquad H=-\frac{d^2}{dx^2}+V(x)b,
 \qquad x \in {\cal
 C}(s) \subset \mathbb{C} ,\qquad
 s \in \mathbb{R} .
 \label{equat}
\end{gather}
For any practical (e.g., phenomenological) purposes one then has the
full freedom of making use of the analyticity properties of $V(x)$
and $\psi(x)$. Strictly in the spirit of~\cite{BB} (or even of
the older futuristic remark~\cite{BT}) one may really convert {\em
the same} dif\/ferential equation~(\ref{equat}) into {\em many}
non-equivalent spectral problems by the mere variation of the
underlying (classes of) curves ${\cal C}(s) \in {\cal R}$.

From such an observation one has to extract the following two
messages. First, one has to accept the idea that the information
about dynamics may be carried not only by the dif\/ferential
expressions~$H$ but also by the topological properties of the
curves ${\cal C}(s)$. Second, once we lost the identif\/ication of~$x$
with the position, we have to reopen the problem of the concrete
realization of the preparation of the system at some initial time
$t=0$ as well as of the specif\/ication of the correct physical
content of the measurement over the system at some subsequent, f\/inal
time $t=T>0$.

Both these questions were already addressed, in the context of
nuclear physics, by Scholtz et al.~\cite{Geyer}. In the similar,
albeit more realistic physical setting these authors immediately
realized that whenever one has, in the usual mathematical sense, $H
\neq H^\dagger$, one has to leave the corresponding ``usual''
Hilbert-space representation $\mathbb{L}^2$ of the states (based on
the use of the ``usual'' inner product $\br \phi|\psi\kt =
\int_\mathbb{R} \phi^*(x) \psi(x) dx$) by a non-equivalent,
adapted, ``standard'' Hilbert-space representation denoted, say, by
the symbol ${\cal H}^{(S)}$ according to our recent brief review~\cite{SIGMA}.

In the spirit of the abstract quantum theory the use of the latter
space ${\cal H}^{(S)}$ returns us to the textbook quantum theory. In
particular, this space must render our given Hamiltonian
self-adjoint, i.e., it must be endowed by an amended,
``metric-dependent'' inner product, $\br \phi|\psi\kt \rightarrow
\br \phi|\Theta|\psi\kt$. The Hilbert-space-metric operator $\Theta$
itself must satisfy a certain set of natural mathematical
requirements of course~\cite{Geyer}.

Starting from the very early applications of the latter idea people
often tried to avoid the direct use of the correct physical space
${\cal H}^{(S)}$ endowed with the nontrivial metric $\Theta$. Indeed
the work within this space requires not only a~suitable modif\/ication
of the related Dirac-notation conventions (in this respect we shall
follow here the recommendations given in~\cite{SIGMA}) but also
a~mind-boggling necessity of using the following ``corrected"
def\/inition of the Hermitian conjugation:
 \begin{gather}
 {\cal T}: \ \ |\psi\kt \ \to \ \bbr \psi| := \br \psi| \Theta .
 \label{adden}
\end{gather}
A comparatively easy insight in this rule is obtained when one
factorizes $\Theta = \Omega^\dagger \Omega$ and replaces the
physical space ${\cal H}^{(S)}$ of kets $|\psi\kt$ and (doubled)
bras $\bbr \psi|$ by its unitarily equivalent ``partner" Hilbert
space ${\cal H}^{(P)}$ of kets
 \begin{gather*}
 |\psi^{(P)}\kt:=|\psi\pkt:=\Omega |\psi\kt
\end{gather*}
and of bras
\begin{gather*}
 \br \psi^{(P)}|:=\pbr \psi|:=\bbr \psi| \Omega^{-1} .
\end{gather*}
On a purely abstract level of thinking one might appreciate that
inside ${\cal H}^{(P)}$ the metric is trivial again,
$\Theta^{(P)}=I$. At the same time the price for simplif\/ication is
too high in general since the Hamiltonian itself acquires a
prohibitively complicated transformed form $\mathfrak{h} =
\Omega H \Omega^{-1}$.

We may summarize that in the majority of applications of models with
apparent non-Hermiticity rule $H \neq H^\dagger$ valid inside the
``friendly'' but ``false'' space  $\mathbb{L}^2 \equiv {\cal
H}^{(F)}$, the concrete form of the above mentioned ``non-unitary
Fourier''~\cite{BG} or ``Dyson's''~\cite{Geyer} mapping $\Omega \neq
(\Omega^\dagger)^{-1}$ is both ambiguous and virtually impossible to
construct. A persuasive conf\/irmation of this generic rule may be
found in~\cite{cubic} where, up to the third order perturbation
corrections,  {\em all} of the eligible operators $\Omega$ were
constructed for the  ``small''  imaginary cubic potential $V(x) =
\lambda {\rm i}x^3$.

It is worth adding that the ambiguity of the ``admissible" metric
operators $\Theta=\Theta(H)$ represents in fact a very unpleasant
feature of the theory. For this reason, its entirely general
factorization $\Theta=\Omega^\dagger\,\Omega$ in terms of the so
called Dyson's mappings $\Omega$ \cite{Geyer} is quite often being
reduced and replaced by its more specif\/ic and less general
forms\footnote{To name just a few let us recollect the parity-using
factorizations $\Theta={\cal CP}$ (= the most popular ansatz using a
``charge" ${\cal C}$ such that ${\cal C}^2=I$, cf.~\cite{BBJ}) and
$\Theta={\cal PQ}$ (where $Q$ denotes quasi-parity, cf.~\cite{pseudo})~-- or simply the square-root factorization
$\Theta=\varrho^2$ with a special, self-adjoint
$\varrho=\varrho^\dagger$~-- cf.~\cite{ali} or~\cite{Jones}.}.

\section{The three-Hilbert-space formulation\\ of quantum mechanics
for quantum toboggans}\label{appendixB}

On the overall background of the three-Hilbert-space formulation of
quantum mechanics as given in \cite{SIGMA} we should emphasize, at
the very beginning, that there exist many important technical
details of its presentation which will vary with the assumptions
concerning the spectral properties of the operators in question. For
this reason let us reduce the corresponding rigorous discussion
to the necessary minimum and assume that all of the operators of our
present interest possess just the point spectrum (composed of a
f\/inite or countably inf\/inite number of eigenvalues and bounded from
below) and  no continuous or residual spectrum.

This assumption will certainly help us to simplify also the language
and to use just the ``minimally'' modif\/ied version of the standard
Dirac's notation as described in more detail in~\cite{SIGMA}. In
particular, let us recollect that the triple possibility of
representation of the ket vectors may be summarized in the following, manifestly time-dependent picture inspired by the very f\/irst proposal and application~\cite{Geyer} of the whole, and very original, scheme in nuclear
physics,
$$
  \begin{array}{c}
    \begin{array}{|c|}
 \hline
  \ {\rm physical} \
 {\cal H}^{\rm (P)}\ (fermions)\\
    {\rm dif\/f\/icult}\ |\psi(t)\pkt\,\equiv\,\Omega(t)\, |\psi(t)\kt
  {\rm } \ \\
 \hline
 \end{array}
\\ {\rm Dyson's \ map}\
 \Omega(t) \    \nearrow \ \  \  \ \ \ \ \ \ \ \
 \ \ \ \ \ \
\ \ \ \ \   \searrow  \ \ {\rm map}\  \Omega^{-1}(t)\\
 \begin{array}{|c|}
 \hline
{\rm friendlier\ }
 {\cal H}^{\rm (F)}\ (bosons) \\
   |\psi(t)\kt
  {\rm =\ computable}\  \\
  \hline
 \end{array}\
 \stackrel{{\rm map}\ \Omega  \Omega^{-1}=I}{ \longrightarrow }
 \
 \begin{array}{|c|}
 \hline
 {\rm \ standard}\
 {\cal H}^{\rm (S)} \ (bosons)\\
   |\psi(t)\kt
  {\rm \ the \ same}\\
 \hline
 \end{array}
\\
\\
\end{array}
$$
In this context the key innovation was twofold. Firstly, the
Fourier-like mapping $\Omega$ (proposed, presumably, by Dyson) was
allowed to be {\em non-unitary}. Secondly, the following
arramengement of the triplet of related, Hilbert-space-dependent bra
vectors proved rather unusual, showing
$$
  \begin{array}{c}
    \begin{array}{|c|}
 \hline%
 {\rm physical}\  \pbr \psi(t)|\,\in\,
 \left ({\cal H}^{\rm (P)}\right )'\
 \\
   \ ={\rm prohibitively\ complicated } \ \\
 \hline
 \end{array}
\\  {\rm map}\
 \Omega^\dagger(t) \ \ \ \
 \nearrow \ \  \  \ \ \ \ \ \
 \ \ \ \ \ \ \ \
\ \ \ \ \  \searrow \ \ \  {\rm map}\ \Omega(t) \\
 \begin{array}{|c|}
 \hline%
  \br \psi(t)|\,\in\,
 \left ({\cal H}^{\rm (F)}\right )'\ \\
 \
  ={\rm  inconsistent\ physics}\    \\%
  \hline
 \end{array}\
 \stackrel{ {\rm map}\ \Theta(t)=\Omega^\dagger\Omega\neq I }{ \longrightarrow }
 \
 \begin{array}{|c|}
 \hline%
 \bbr \psi(t)|\,\in\,
 \left ({\cal H}^{\rm (S)}\right )'\
 \\
  ={\rm  innovated \ conjugation}  \\%
 \hline
 \end{array}
\\
\\
\end{array}
$$
Two conclusions may be deduced.
Firstly, the mapping ${\cal T}:\br \psi | \rightarrow  \bbr \psi
|
 =\br \psi |\Theta$
may be perceived as a ``very natural and very physical'' or
``amended" Hermitian conjugation (reason: the Hamiltonian s made
self-adjoint in ${\cal H}^{\rm (S)}$). Secondly, the ``double-bra''
vectors may be constructed as eigenvectors of the
no-metric-conjugate $H^\dagger$ (which dif\/fers from $H$ itself).
Then, one can reconstruct the metric using the explicit formula
\begin{gather*}
  \Theta=\sum  |\psi\kkt \bbr
 \psi| .
\end{gather*}

\subsection[The hermitized ``generalized eigenvalue problems'' ($W \neq I$)]{The hermitized ``generalized eigenvalue problems'' ($\boldsymbol{W \neq I}$)} \label{uctvrta}

Inside the constructively inaccessible ``paternal'', physical Hilbert
space ${\cal H}^{\rm (P)}$ our Sturm--Schr\"{o}\-din\-ger equation is
self-adjont,
 \begin{gather*}
 h^{}\,|\,{\lambda}\pkt
 =\lambda\,w\,|\,{\lambda}\pkt .
\end{gather*}
Inside the same space we may work with the
 Sturmianic orthogonality
relations
 \begin{gather*}
 \pbr \lambda\,|\,w\,|\,\lambda'\pkt\ =\
 \pbr \lambda\,|\,w\,|\,\lambda\pkt\,\cdot\,
 \delta_{\lambda,\lambda'}
\end{gather*}
and with the Sturmianic completeness relations,
 \begin{gather*}
 I = \sum_{\lambda}\,|\,\lambda\pkt
 \,
 \frac{1}{\pbr \lambda\,|\,w\,|\,\lambda\pkt}
 \,\pbr \lambda\,|\,w .
\end{gather*}
In the same manner there exists the following Sturmianic spectral
representation of the Hamiltonian,
 \begin{gather*}
 h = \sum_{\lambda}\,w\,|\,\lambda\pkt\,
 \frac{\lambda}{\pbr \lambda\,|\,w\,|\,\lambda\pkt}
 \,\pbr \lambda\,|\,w .
 \end{gather*}

\subsection[Back to ${\cal H}^{(F)}$]{Back to $\boldsymbol{{\cal H}^{(F)}}$}

As long as the manifestly unphysical and ``false'' Hilbert space
${\cal H}^{\rm (F)}$ is still, by assumption, maximally ``friendly''
to computations, it makes sense to translate the above-listed
formulae to this space as well. Thus, one arrives at the
orthogonality relations in ${\cal H}^{(F)}$,
 \begin{gather*}
 \langle\,\lambda\,|\,
 \Omega^\dagger w \Omega\,|\,\lambda'\,\kt =
 \langle\,\lambda\,|\,\Theta W
 \,|\,\lambda'\,\kt =
 \langle\,\lambda\,|\,\Theta W
 \,|\,\lambda\,\kt\,\cdot\,
 \delta_{\lambda,\lambda'}
 \end{gather*}
as well as at the following completeness relations inside ${\cal
H}^{(F)}$,
 \begin{gather*}
 I = \sum_{\lambda}\,|\,\lambda\,\kt \,
 \frac{1}{
 \langle\,\lambda\,|\,\Theta W
 \,|\,\lambda\,\kt}
 \,
 \langle\,\lambda\,|\,
 \Theta W .
\end{gather*}
Thirdly, the spectral decomposition of the Hamiltonian in ${\cal
H}^{(F)}$ has the following form,
 \begin{gather*}
 H = \sum_{\lambda}\,W\,|\,\lambda\,\kt
 \,
 \frac{\lambda}{
 \langle\,\lambda\,|\,\Theta W
 \,|\,\lambda\,\kt}
 \,\langle\,\lambda\,|\,
 \Theta W .
\end{gather*}
Although the choice of the convention $
|\,\lambda\,\rangle\!\rangle=\Theta|\,\lambda\,\rangle$ represents
just one of many alternative possibilities, its preference also
simplif\/ies the normalization of the overlap
$\langle\,\lambda\,|\,\Theta W \,|\,\lambda\,\kt=1$.

\subsection[$\Theta W$ stays positive definite]{$\boldsymbol{\Theta W}$ stays positive def\/inite}

It is necessary to keep in mind that the use of the formal
abbreviations
 \begin{gather*}
 |\,\psi\,\} = W\,|\,\psi\,\kt ,
 \qquad
 |\,\psi\,\ktz = W^\dagger\,|\,\psi\,\kkt
 \end{gather*}
may perceivably simplify our formulae. First, it leads to an
 amended orthogonality,
 \begin{gather*}
 \langle\,\lambda\,|\,\Theta W
 \,|\,\lambda'\,\kt=
 \brz\,\lambda\,|
 \,\lambda'\,\kt=
 \bbr\,\lambda\,|\,\lambda'\,\}=
 \delta_{\lambda,\lambda'}
 .
\end{gather*}
Second, the alternative completeness relations read
 \begin{gather*}
 I = \sum_{\lambda}\,|\,\lambda\,\kt \,
  \brz\,\lambda\,| = \sum_{\lambda}\,|\,\lambda\,\} \,
  \bbr\,\lambda\,|
 .
\end{gather*}
Third, the
 spectral-representation expansions
 \begin{gather*}
 W = \sum_{\lambda}\,|\,\lambda\,\} \,
  \brz\,\lambda\,|
 ,\qquad
 H = \sum_{\lambda}\,|\,\lambda\,\}
 \,
 {\lambda}
 \,\brz\,\lambda\,|
\end{gather*}
represent, this time, not only the Hamiltonians but also the
weight-operators.

In a f\/inal step of our review of the formalism we may rewrite the
Dieudonn\'{e} equations in the form
 \begin{gather*}
 \sum_{\lambda}\,|\,\lambda\,\ktz \,\lambda\,\{\lambda|\,\Theta=
  \sum_{\lambda}\,\Theta\,|\,\lambda\,\} \,\lambda\,\brz
  \lambda|
\end{gather*}
or
 \begin{gather*}
 \sum_{\lambda}\,|\,\lambda\,\ktz \,\{\lambda|\,\Theta=
  \sum_{\lambda}\,\Theta\,|\,\lambda\,\} \,\brz
  \lambda|\,
\end{gather*}
and easily derive
 the f\/inal single-series spectral formula
 \begin{gather*}
 \Theta=  \sum_{\lambda}\,|\,\lambda\,\kkt \,
  \brz\,\lambda\,|
\end{gather*}
for the metric \cite{Hendrik}.

\section[Imaginary cubic toboggans in the large-$L$ perturbation regime]{Imaginary cubic toboggans\\ in the large-$\boldsymbol{L}$ perturbation regime}\label{appendixC}

For readers interested in the practical and phenomenological aspects
of our present considerations let us brief\/ly recall the results of
calculations~\cite{anom} where the popular~\cite{cubic}
non-Hermitian potential $V(x)=ix^3$ has been studied in the QTM
kinematical regime {\em and} in the high-angular-momentum dynamical
domain.

First, the rectif\/ication change of variables did lead to the
Schr\"{o}dinger (or rather Sturm--Schr\"{o}dinger) bound-state
problem
 \begin{gather*}
 \left [-\frac{d^2}{dy^2}
 +\frac{L(L+1)}{y^2}
 +{\rm
 i} (-1)^N (2N+1)^2 y^{10 N+3}
 \right ] \psi_n(y)
 =
(2N+1)^2
 y^{4N} E^{[N]}_n \psi_n(y),
\end{gather*}
where $N$ is a small integer while $L$ is only assumed real and
(very) large. We are returning to these results because they very
well illustrate not only the method (of asymptotic estimates) but
also the comparatively friendly nature of the spectral consequences
of the topological non-triviality of the QTM integration paths
${\cal C} = {\cal C}^{(N)}$ where, roughly speaking, the small and
positive integer $N$ is a winding number of the tobogganic path.

\subsection[The method:  $1/L$ series at $N=0$ ($L=\ell$, $y = q^{(0)}$)]{The method:  $\boldsymbol{1/L}$ series
at $\boldsymbol{N=0}$ ($\boldsymbol{L=\ell}$, $\boldsymbol{y = q^{(0)}}$)}

Near a (local as well as global) minimum $V_{\rm ef\/f}(Q)$ of any
suf\/f\/iciently smooth function $V_{\rm ef\/f}(q)$ we may often work just
with the truncated forms of the Taylor series
\begin{gather*}
  V_{\rm ef\/f}(q) =    V_{\rm ef\/f}(Q) +  \frac{1}{2} V_{\rm ef\/f}''(q)\,\xi^2
     +  \frac{1}{6} V_{\rm ef\/f}'''(q) \xi^3+
   \cdots
,
\end{gather*}
where $\xi = q-Q$ and where we showed in~\cite{anom} that for
the imaginary cubic oscillators one may select $2
\ell(\ell+1)=3{\rm i}Q^5$ = large and restrict the set of all of the
available (complex) roots $Q=Q_j$, $j=1,2,\ldots,5$  just to the
$j=1$ preferred item. Then we deduced

\begin{lemma} In the non-tobogganic case the low-lying imaginary-cubic spectrum is
prescribed by the estimate
\begin{gather*}
 E_n^{[0]}
  =-\frac{5 \tau^{3}}{2}
  +\sqrt{\frac{15 \tau}{2}} (2n+1)
 +
    {\cal O}\big(\tau^{-3/4}\big) ,\qquad n = 1, 2, \ldots,
\end{gather*}
where
 \begin{gather*}
     \tau=
      \left |
  \left (
    {\frac{2}{3} {\ell(\ell+1)}}
    \right )^{1/5}
 \right |
 \qquad \text{in} \quad Q_j= -{\rm i} \tau \exp\left(\frac{2{\rm i}\pi (j-1)}{5}\right),
\qquad j = 1,2,\ldots,5.
 \end{gather*}
\end{lemma}

\begin{proof}
With $j=1$ and with a  small $\sigma = 1/\tau^{1/4}$ we may
rescale $\xi \to \sigma \xi$ and re-write
 \begin{gather*}
  V_{\rm ef\/f}(q) =   -\frac{5}{2} \tau^3 + \frac{15  \tau}{2} \xi^2
   - 5 {\rm i} \xi^3  +
    {\cal O}\big(\tau^{-1}\big)
    \label{tayl}
  \end{gather*}
leading to the leading-order harmonic-oscillator Schr\"{o}dinger
equation
 \begin{gather*}
 \left [-\frac{d^2}{d\xi^2} -\frac{5}{2\,\sigma^{10}}
 + \frac{15}{2} \xi^2
   - 5 \sigma^5  {\rm i} \xi^3  +
    {\cal O}\big(\sigma^6\big)
 \right ] \varphi^{[0]}(-{\rm i}\tau + \xi)
 =
 \sigma^2  E_n^{[0]}
 \varphi^{[0]}(-{\rm i}\tau + \xi) .\tag*{\qed}
 \end{gather*}
 \renewcommand{\qed}{}
 \end{proof}

\subsection[The conclusion: the spectrum {\em is} $N$-dependent]{The conclusion: the spectrum {\em is} $\boldsymbol{N}$-dependent}

Once we introduce $T=T_j(L)$ as a set of roots of the
elementary algebraic equation
 \begin{gather*}
 2L(L+1)=(2N+1)^2(10N+3) T^{10N+5}
\end{gather*}
we may abbreviate
\begin{gather*}
     \tau=\tau^{(N)}=
      \left |
  \left (
    \frac{2\,L\,(L+1)}
    {(2N+1)^2(10N+3)}
    \right )^{1/(10N+5)}
 \right |
 \end{gather*}
and specify all the solutions,
 \begin{gather*}
 T_1=-{\rm i}\tau , \qquad
 T_{j}=T_{j-1} e^{2{\rm i} \pi/(10N+5)  } ,\qquad
 j=2, 3,
 \ldots, 10N+5 .
 \end{gather*}
We shall only use here  $T_1=-{\rm i}\tau$ giving the maximal size
of the complex shift $\varepsilon = \tau$.

\begin{theorem} In the tobogganic cases numbered by the winding number~$N$ the low-lying ima\-gi\-na\-ry-cubic spectrum is
prescribed by the estimate
 \begin{gather*}
 E_n^{[N]}
  =-\frac{10N+5}{2}  \tau^{6N+3}
  +
\frac{2n+1}{2N+1}
    \sqrt{\frac{(10N+3)(10N+5)}
    {2}
    } \tau^{N+1/2}
 +
    {\cal O}\big(\tau^{-(6N+3)/4}\big) .
\end{gather*}
\end{theorem}

\begin{proof}
The use of the standard methods gives
 \begin{gather*}
  V_{\rm ef\/f}(y) =  -\frac{1}{2} (2N+1)^2(10N+5) \tau^{10N+3} +
  \omega^2_{(N)} \tau^{10N+1} \xi^2+
 \mu(N) \tau^{10N} \xi^3+
    {\cal O}\big(\tau^{10N-1}\big)
\end{gather*}
with
 \begin{gather*}
    \omega_{(N)}=(2N+1)
    \sqrt{\frac{(10N+3)(10N+5)}
    {2}
    } .
\end{gather*}
The resulting approximate equation
\begin{gather*}
 \left [-\frac{d^2}{d\xi^2}
    +
  \omega^2_{(N)}\,\tau^{10N+1} \xi^2  + \mu(N) \tau^{10N} \xi^3+\cdots
  \right ] \psi_n(-{\rm i}\tau + \xi)
\\
\qquad{}
  = \left \{
  E^{[N]}_n (2N+1)^2
 \left[ \tau^{4N}
 +\ldots\right ]
  +\frac{1}{2} (2N+1)^2(10N+5) \tau^{10N+3}
 \right \}
 \psi_n(-{\rm i}\tau + \xi)
\end{gather*}
rescales in the form $\xi \to \sigma \xi$ using $\sigma =
\tau^{-(10N+1)/4}$,
 \begin{gather*}
 \left [-\frac{d^2}{d\xi^2} -\frac{5}{2 \sigma^{10}}
 + \frac{15}{2} \xi^2
   - 5 \sigma^5  {\rm i} \xi^3  +
    {\cal O}\big(\sigma^6\big)
 \right ] \varphi^{[N]}(-{\rm i}\tau + \xi)
 =
  \sigma^2 E_n^{[N]}\varphi^{[N]}(-{\rm i}\tau + \xi)
\end{gather*}
and the closed formula for the energies follows.
 \end{proof}

 \begin{figure}[t]
\centering
\includegraphics[angle=270,width=75mm]{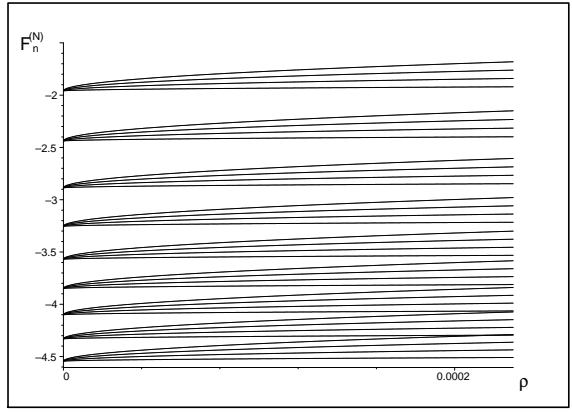}

\caption{The rescaled tobogganic energies $
F_n^{(N)}=\rho^{3/5}\,E_n^{(N)}(\rho)$ in the imaginary cubic well.
The f\/irst four lowest energies with $n = 0, 1, 2, 3$ are sampled at
very small $\rho = 1/(\ell+1/2)^2$. The increase of the winding $N =
0, 1, \ldots, 8$ pushes the spectrum downwards. \label{fionset}}
\end{figure}

\begin{figure}[t]
\centering
\includegraphics[angle=270,width=75mm]{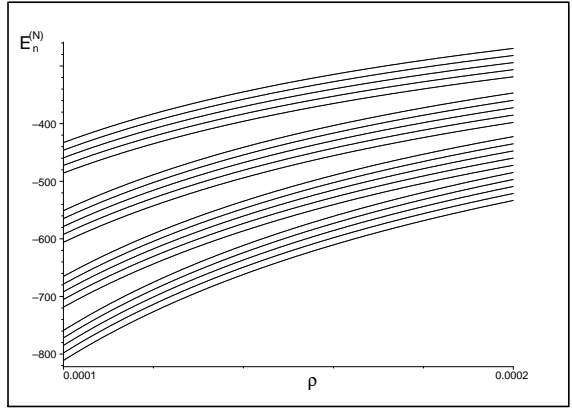}

 \caption{The unrescaled tobogganic energies $
E_n^{(N)}(\rho)$ in the imaginary cubic well. The f\/irst f\/ive lowest
energies with $n = 0, 1, 2, 3,4$ are sampled in a non-asymptotic
interval of $\rho >\rho_0>0$. The result is shown for windings $N =
0, 1, 2 $ and $3$. \label{onset}}
\end{figure}

In terms of a parameter $\rho = 1/(\ell+1/2)^2$ our approximate
spectrum may be also rescaled giving the function $
F_n^{(N)}=\rho^{3/5} E_n^{(N)}$ with the smooth and
$\rho$-dependence which remains bounded even in the
inf\/initely-large-$\ell$ limit. This is illustrated by
Fig.~\ref{fionset} (units omitted as inessential).

Naturally, the scaling is unnecessary for the f\/inite values of
$\ell$. Our last Fig.~\ref{onset} displays, therefore, the
segments of the spectra of energies (with $n=0,1,2,3,4$)  for the
f\/irst four winding numbers $N=0,1,2$ and $3$.

\subsection*{Acknowledgements}

Work supported by M\v{S}MT ``Doppler Institute" project Nr. LC06002,
by the GA\v{C}R grant Nr. P203/11/1433, by the Institutional
Research Plan AV0Z10480505 and, last but not least, by the
hospitality of STIAS in Stellenbosch in November 2010.

\pdfbookmark[1]{References}{ref}
\LastPageEnding

\end{document}